\documentclass{nature_max}

\usepackage{graphicx}
\usepackage{amsmath}
\usepackage{amsfonts}
\usepackage{bm}
\usepackage{color}
 
\title{A monolayer transition metal dichalcogenide as a 
topological excitonic insulator}

\author{Daniele Varsano$^1$, Maurizia Palummo$^2$, Elisa Molinari$^{1,3}$
        \& Massimo Rontani$^1$}

\begin{document}

\maketitle

\begin{affiliations}
 \item CNR-NANO, Via Campi 213a, 41125 Modena, Italy.
 \item INFN, Dipartimento di Fisica, Universit{\`a} degli Studi di Roma Tor Vergata, 
	 Via Della Ricerca Scientifica 1, 00133 Roma, Italy.
 \item Dipartimento di Scienze Fisiche, Informatiche e Matematiche, 
       Universit{\`a} degli Studi di Modena e Reggio Emilia, 41125 Modena, 
       Italy.
\end{affiliations}

\begin{abstract}
Monolayer transition metal dichalcogenides in the $T'$ phase promise 
	to realize the quantum spin Hall (QSH) 
	effect\cite{Kane2005a} at room temperature, 
because they exhibit a prominent spin-orbit gap between inverted  
bands in the bulk\cite{qian2014quantum,Jarillo-Herrero2018}. 
Here we show that the binding energy of electron-hole
pairs excited through this gap is larger than the gap itself in MoS$_2$, a paradigmatic material that we investigate from first principles by many-body perturbation theory\cite{Onida-Reining-Rubio_2002} (MBPT).
This paradoxical result hints at the instability of the $T'$ phase against the spontaneous generation of excitons, and indeed we find  that it gives rise to a recostructed `excitonic insulator' ground state\cite{Keldysh1964,Cloizeaux1965,Kohn1967,Halperin1968,Volkov1973}.
Importantly, we show that in this system topological and excitonic
order cooperatively enhance the bulk gap by breaking 
the crystal inversion symmetry, 
in contrast to the case of bilayers\cite{Bernevig2006,Konig2007,Liu2008,Knez2011,Budich2014,Pikulin2014,Du2017,Xue2018,Zhu2019} where the frustration between the two orders is relieved by breaking time reversal 
symmetry\cite{Pikulin2014,Xue2018,Zhu2019}.
The excitonic topological insulator departs distinctively from the 
bare topological phase as it lifts the band spin degeneracy, 
which results in circular dichroism.
A moderate biaxial strain applied to the system leads 
to two additional excitonic phases, different in their topological 
character but both ferroelectric\cite{Portengen1996b,fei-cobden_2018} as  
an effect of electron-electron interactions.
\end{abstract}


The monolayer transition metal dichalcogenides
that were recently proposed as candidates for the QSH effect
all have overlapping metal-$d$ conduction and 
chalcogenide-$p$ valence bands\cite{qian2014quantum}. 
Such `band inversion' makes the system
either a narrow-gap semiconductor, due to $p-d$ 
spin-orbit hybridization (Fig.~1c), or a semimetal, whose band edges
are displaced in momentum space. In both cases long-range
Coulomb attraction, which is poorly screened in two dimensions, 
tends to bind electrons ($e$) at the bottom of conduction 
band with holes ($h$) at
the top of valence band, thus giving rise to excitons. 
If the {\it e-h} binding energy is larger than the semiconductor gap 
(or if it is non vanishing in the semimetal), 
then excitons will spontaneoulsy form 
and condense, 
until a correlated gapped phase is 
built at thermodynamic equilibrium: the excitonic insulator proposed in the sixties\cite{Keldysh1964,Cloizeaux1965,Kohn1967,Halperin1968}. 
This paradigm has been recently invoked for the  QSH insulator WTe$_2$,
as its bulk gap is strongly sensitive to 
temperature\cite{Jia2017,Fei2017,Song2018} and
doping\cite{Sajadi2018}, 
whereas in the absence of excitonic effects MBPT predicts 
semimetallic behaviour\cite{qian2014quantum}. 
Mounting evidence of the excitonic insulator 
has been accumulating in the last two years 
in low-dimensional 
materials\cite{Varsano-Rontani2017,Dean2017,Kim2017,Du2017}---noticeably 
transition metal 
dichalcogenides\cite{Kogar-Abbamonte_2017,Kono2017,Kaiser2018}.

The relation between topological and excitonic order is intriguing,
as the former emerges in a noninteracting picture whereas the latter 
is driven by {\it e-h} interactions. So far the problem has been 
discussed for bilayers\cite{Budich2014,Pikulin2014,Du2017,Xue2018,Zhu2019}: 
the spin-orbit gap, associated with topological
order, depends on interlayer tunneling but the exciton 
binding does not, as $e$ and $h$ in separate layers 
remain bound by long-range attraction. 
This leads to scenarioes of frustration between 
the QSH phase and the topologically trivial excitonic phase. 
The monolayer case is different, since screening is suppressed as the gap opens: here both spin-orbit gap and exciton binding 
are affected by interband hybridization.  

Here we study an archetypical member of the $T'$ family of QSH candidates, monolayer MoS$_2$, through first-principles calculations
by means of MBPT, and take into account {\it e-h} binding by solving the Bethe-Salpeter 
equation\cite{Noziers1964}. 
This allows us to assess the inherent excitonic instability of the material.
We then demonstrate the coexistence of topological and excitonic orders
through a self-consistent approach, which
predicts a chiral ground state wave function with unique fingerprints.

\begin{figure}[htbp]
\setlength{\unitlength}{1 cm}
\begin{picture}(16,6)
\put(0.0,-0.5){\includegraphics[trim=0cm 0cm 0 0,clip=true,width=16.0cm]{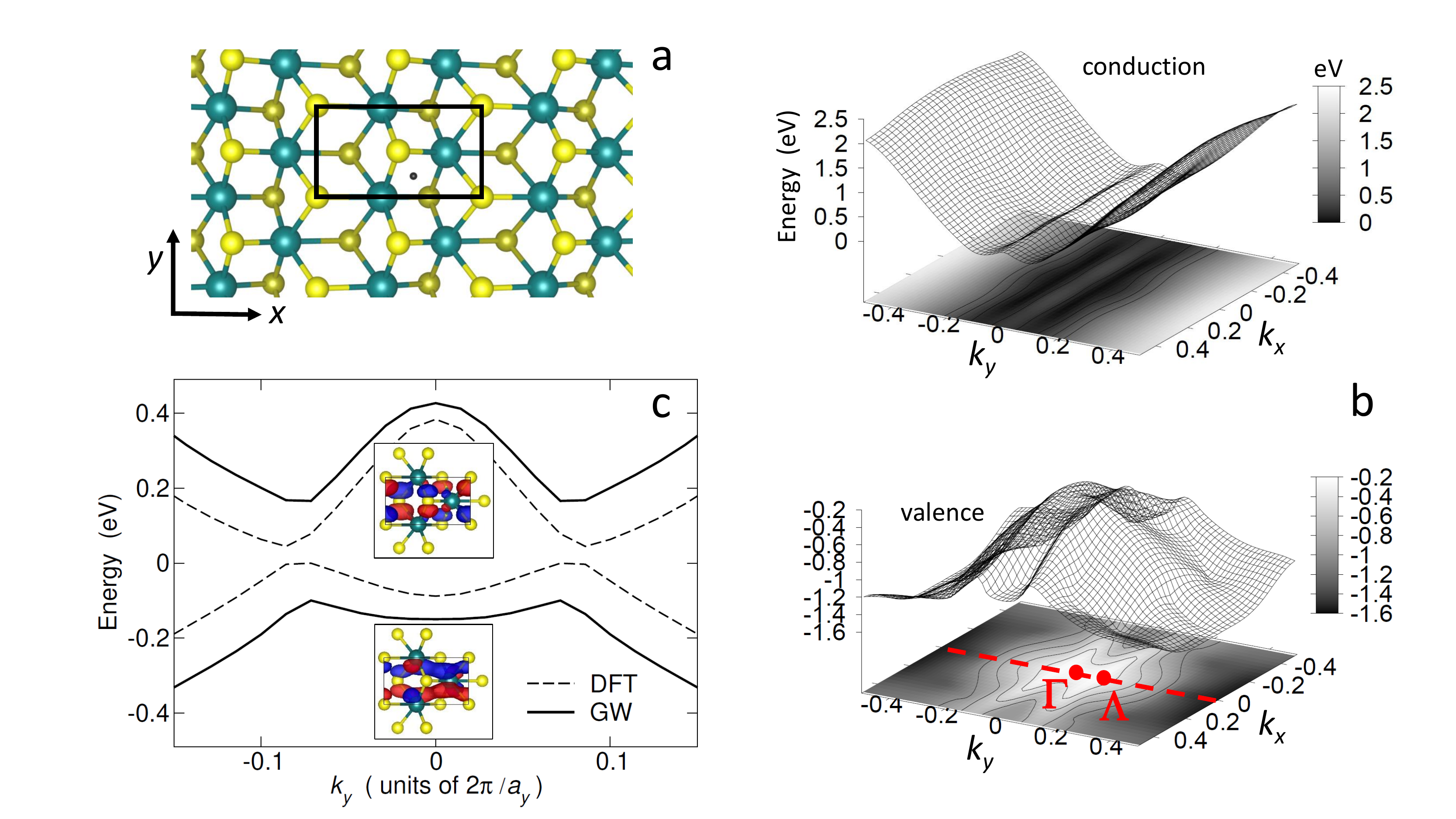}}
\end{picture}
\caption{
{\bf Electronic band structure of T$'$-MoS$_2$. }
{\bf a,} Stick-and-ball	model of the crystal (projection on the plane of Mo atoms). The turquoise (yellow) colour
labels Mo (S) atoms, and the light / dark shadow distinguishes S atoms lying above / below the Mo plane. The black dot highlights
the inversion center, located at 
midpoint between two Mo atoms. The $y$ axis of the 
unitary cell (black frame) is parallel to the Mo zig-zag 
chain.
{\bf b,} Plot of conduction and valence 
energy bands, as obtained from first-principles many-body perturbation theory (GW). The unit of $k_i$   
is $2\pi/a_i$, with $a_i$ being the lattice constant in the $i$th direction, 
$i=x,y$. {\bf c,} Band dispersion
along the $\Gamma-\Lambda$ cut (red line in panel b) from 
density functional theory (DFT, dashed curve) 
and GW (solid curve) calculations,
respectively. Absolute values of GW bands are shifted in energy to facilitate comparison with DFT bands.  
Insets show the isosurfaces of
Bloch state wave functions at the $\Gamma$ point, 
with blue / red colours 
distinguishing negative / positive values.
}
\end{figure}

\section*{Results}

{\bf Exciton binding and instability.}
In spite of the small point symmetry group, 
$T'$-MoS$_2$ has an inversion center, located 
at the midpoint between two neighbour Mo atoms (black dot in Fig.~1a): 
these form a zig-zag chain, parallel to
the $y$ axis, which is characteristic of the $T'$ phase.
We compute the energy bands   
from first principles, 
including spin-orbit interaction at the density functional 
theory (DFT) level, and then evaluating
many-body corrections  
within the GW approximation (Methods).
The resulting band structure is 
highly anisotropic (Fig.~1b), the conduction band being almost
flat in the direction perpendicular to the zig-zag chain. 
The cut of energy surfaces along the $\Gamma\Lambda$ direction 
(dashed red line in the Brillouin zone domain of Fig.~1b)
clarifies why $T'$-MoS$_2$ is a QSH insulator. As shown in
Fig.~1c, the overlap of $p$ and $d$ bands
forms two valleys located at {\bf k} = $\pm \Lambda$, 
similar to Dirac valleys of graphene. 
Like in the Kane-Mele model\cite{Kane2005a}, spin-orbit interaction
opens a gap, acting as a valley-dependent magnetic field for each spin 
species: hence the valley contributions to
the topological invariant $\mathbb{Z}_2$ do not cancel 
out\cite{Kane2005a,qian2014quantum}.
Importantly, the GW renormalization of the DFT gap
is gigantic (respectively solid and dashed lines in Fig.~1c),  
which points to the relevance of electronic interactions.

\begin{figure}[htbp]
\setlength{\unitlength}{1 cm}
\includegraphics[trim=0cm 5cm 0 0,clip=true,width=15.5cm]{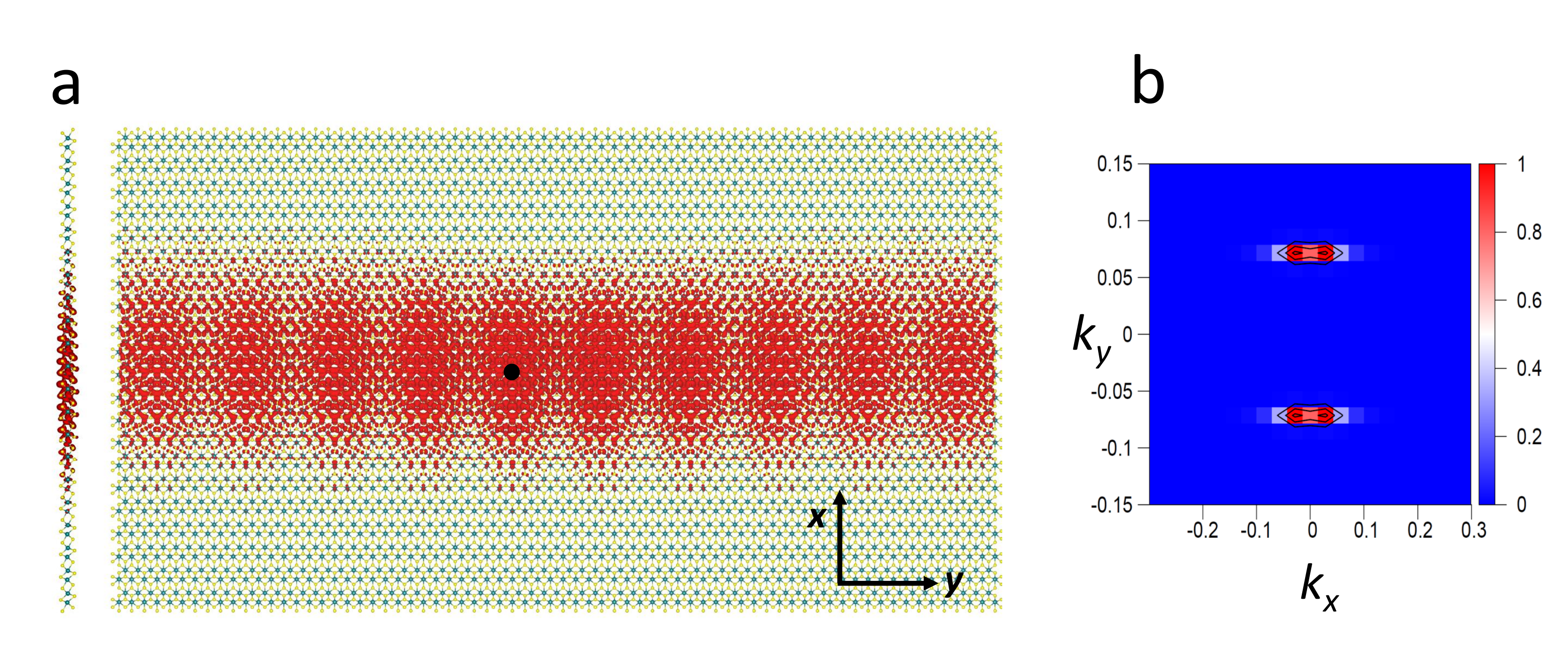}
\caption{
{\bf 
Exciton wave function from first principles.}
{\bf a,} 
Exciton wave function square modulus, 
as obtained from Bethe-Salpeter equation (GW-BSE).
The contour plot (red colour) is the probability density 
to locate the bound electron once the hole position is fixed (black dot).  The figure contains 21 and 69 unit cells in the $x$ and $y$ direction, respectively. Note the delocalization in real space along the $y$ direction of Mo zig-zag chains. 
{\bf b,} Exciton wave function square modulus 
in the reciprocal space region around the $\Lambda$ points. 
}
\end{figure}

From the solution of 
Bethe-Salpeter equation (Methods) we find that the binding energy of
the lowest exciton exceeds the GW gap by 32 meV, hence $T'$-MoS$_2$ is unstable against the spontaneous generation of excitons.
The exciton probability weight 
in momentum space is localized in the two $\Lambda$ valleys (Fig.~2b): the weight is stretched along the
$k_x$ direction, following the energy profile of 
uncorrelated {\it e-h} pairs excited with zero total momentum.
This exciton is eight-fold degenerate within numerical accuracy, 
as multiplet states include
both bonding and antibonding combinations of the wave functions in the two valleys, for all possible $e$ and $h$ spin projections
along the two-fold screw axis $y$ (cf.~Fig.~1a); the spin full rotational symmetry is reduced to a rank-two representation by spin-orbit interaction. 
The exciton wave function in real space is shown in Fig.~2a as the conditional probability  
of finding a bound electron (red colour), provided the hole
position is fixed (black dot): this density is substantially
delocalized in the direction of the Mo zig-zag chain, consistently with Fig.~2b.

\begin{figure}[htbp]
\setlength{\unitlength}{1 cm}
\includegraphics[trim=0cm 0cm 0 0,clip=true,width=16.0cm]{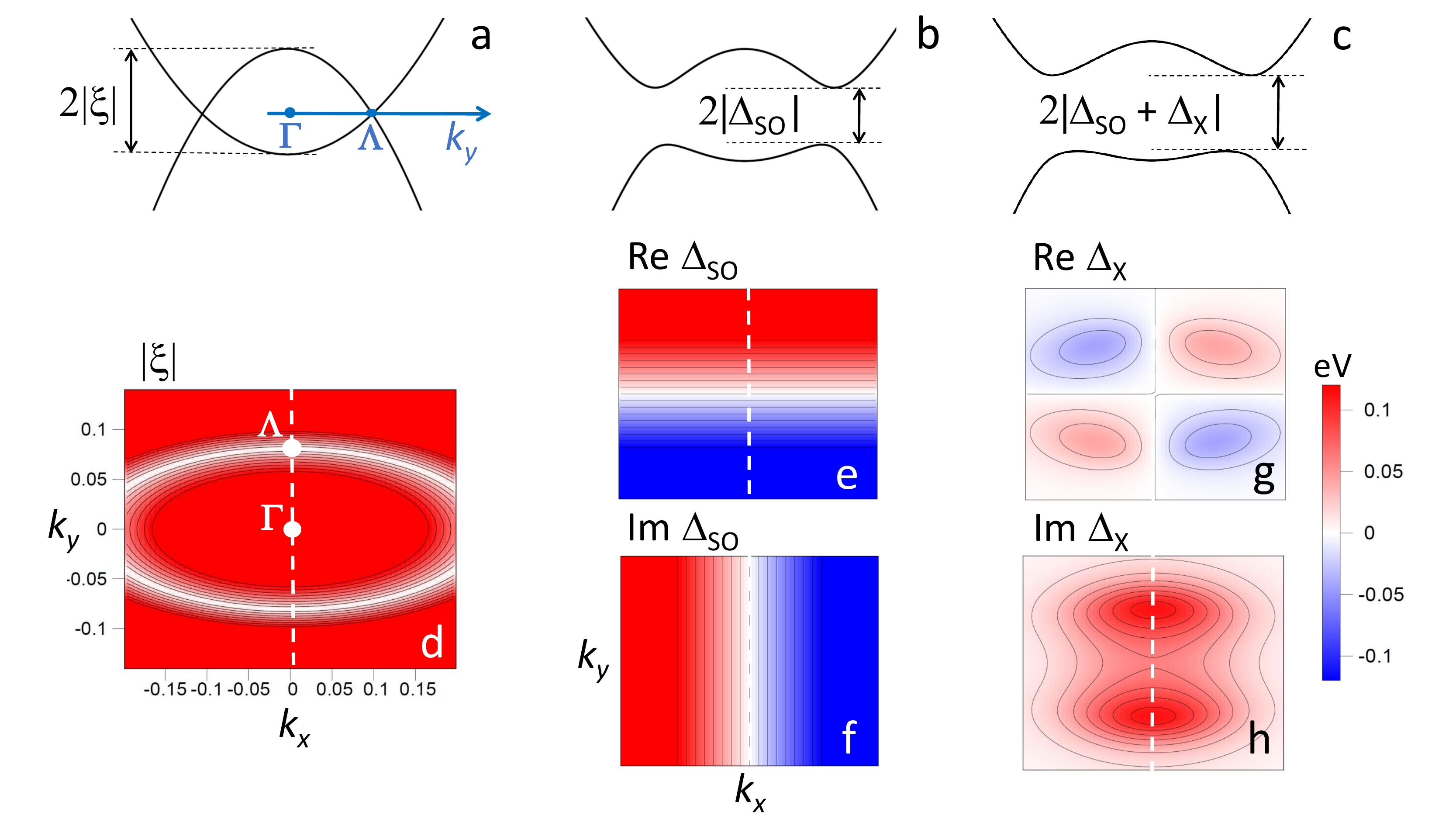}
\caption{
{\bf Topological vs excitonic order.} 
{\bf a-c,} Sketches of gap-opening mechanisms. Starting from the semimetal
in the absence of interband hybridization (panel a), a gap opens 
due to spin-orbit coupling, $\Delta_{\text{SO}}$ (panel b).
The self-consistent excitonic hybridization, $\Delta_{\text{X}}$,
further increases the gap (panel c), which is $\approx 2 \left|\Delta_{\text{SO}} +
\Delta_{\text{X}}\right|$ for the spin 
branch $\lambda = +$ and located close to $\Lambda$. 
{\bf d-h,} Contour maps of 
half $e-h$ excitation energy of the semimetal, $\left|\xi(\text{\bf k}) \right|$ (panel d),
$\Delta_{\text{SO}}$
(real and imaginary parts respectively in panels e and f), 
$\Delta_{\text{X}}$ (panels g and h).  
Since both Im\{$\Delta_{\text{SO}}$\} and 
Re\{$\Delta_{\text{X}}$\} 
vanish around $\Lambda$,
the two remaining gap-opening components, 
Re\{$\Delta_{\text{SO}}$\} and 
Im\{$\Delta_{\text{X}}$\}, add together without interfering. 
}
\end{figure}

\noindent {\bf Topological excitonic insulator.} The excitonic instability reminds us of the Cooper problem of two bound electrons filling a Fermi sea: whereas the gluing of $e-e$ pairs heralds the transition to the Bardeen-Cooper-Schrieffer superconductor, 
the strong binding of $e-h$ pairs signals the formation of the excitonic 
insulator: both Cooper pairs and excitons collectively enforce
the many-body gap, which is the self-consistent order parameter\cite{Kohn1967b}. 
Were $T'$-MoS$_2$ a semimetal (Fig.~3a), the analogy would be 
complete at the formal level, 
as the excitonic order parameter, $\Delta_{\text{X}}$, would then hybridize 
conduction and valence bands, opening a gap\cite{Kohn1967}.
However, the actual bands are effectively hybridized by spin-orbit interaction, 
$\Delta_{\text{SO}}$ (Fig.~3b), hence the role of $\Delta_{\text{X}}$ requires further clarification (Fig.~3c). 

Following the seminal work by Volkov and Kopaev\cite{Volkov1973}, a key observation is that $\Delta_{\text{SO}}$ and $\Delta_{\text{X}}$
have opposite parity in {\bf k} space. In order to preserve the inversion symmetry of the crystal, $\Delta_{\text{SO}}$ must be odd, $\Delta_{\text{SO}}(\text{\bf k}) = 
- \Delta_{\text{SO}}(-\text{\bf k})$,
as the periodic parts of conduction and valence states transform like $p_y$ and $d_{yz}$ orbitals, respectively, as illustrated 
by their wave functions at $\Gamma$ (insets in Fig.~1c). Since
$\Delta_{\text{X}}$ 
is associated to the lowest-exciton wave function 
in {\bf k} space\cite{Kohn1967b}, it must have $s$-wave symmetry,   
$\Delta_{\text{X}}(\text{\bf k}) = 
\Delta_{\text{X}}(-\text{\bf k})$. 
Besides, the spin degeneracy associated with the exciton that drives
the instability rules out the breaking of time reversal symmetry. 
Together, these conditions provide us with a tractable two-band model, 
by reducing the number of independent order parameters from eight 
(the spin-resolved, complex interband hybridizations) to two, i.e., 
the real and imaginary parts of $\Delta_{\text{X}}(\text{\bf k})$ 
(see Methods). 

\begin{figure}[htbp]
	\setlength{\unitlength}{1 cm}
\includegraphics[trim=0cm 0cm 0 0,clip=true,width=16.0cm]{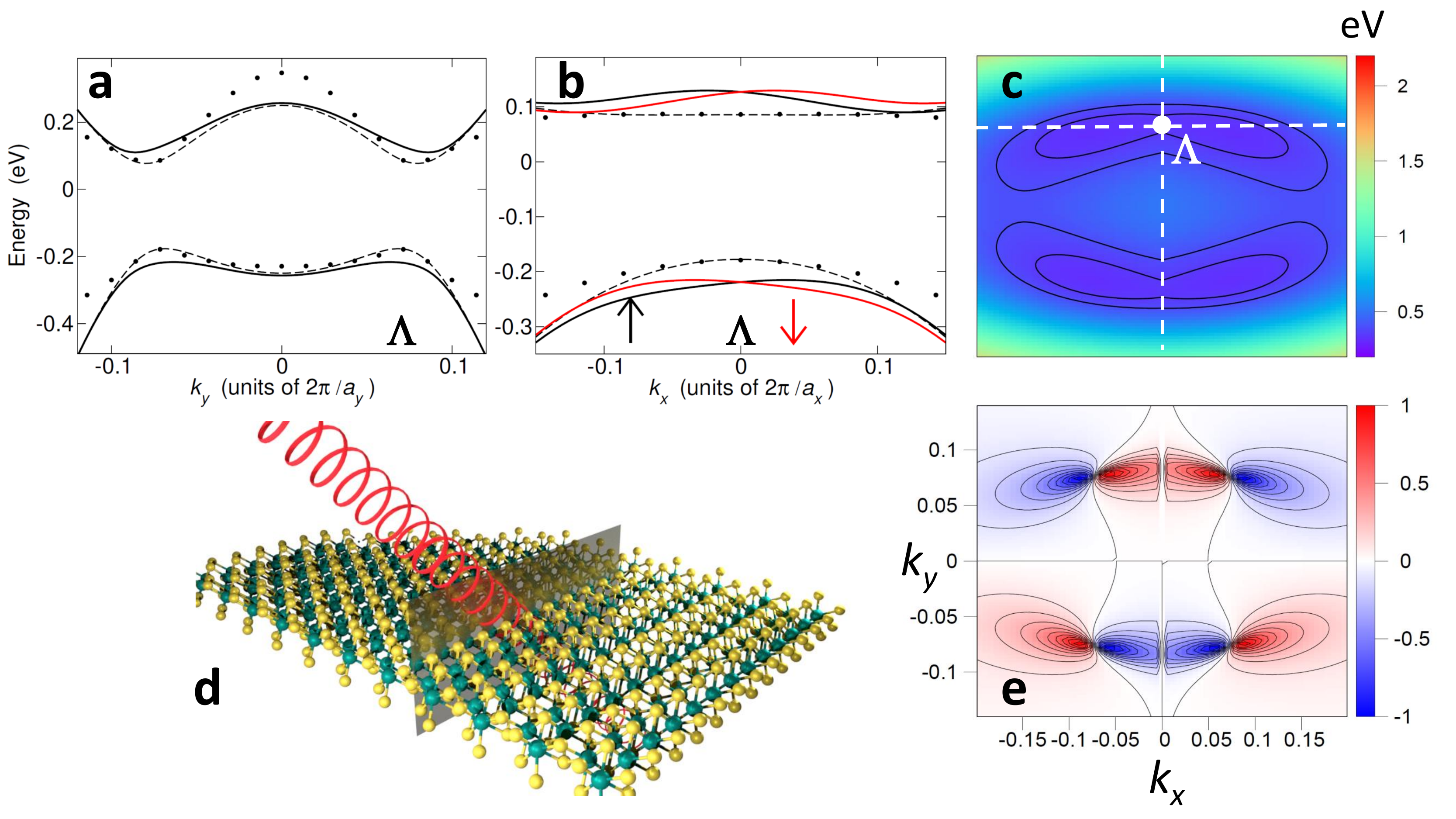}
\caption{
{\bf Signatures of the topological excitonic insulator.}
{\bf a-b,} Energy bands of the topological excitonic insulator
along directions $k_x=0$ 
(panel a, solid line)  
and $k_y=\Lambda$ (panel b, solid black and red lines label opposite spin projections). The two-fold rotational symmetry 
along the $y$ axis 
preserves spin degeneracy for $k_x=0$.
Dotted lines are first-principles GW data and dashed lines are 
effective-mass fits for $\Delta_{\text{X}}=0$. 
{\bf c,} Contour map of lowest $e-h$ excitation
energy in {\bf k} space, $E_c$({\bf k}) $-E_v$({\bf k}).
White dashed lines are the cuts shown in panels a and b.
{\bf d,} Optical absorption configuration: the wave vectors of incoming circularly polarized photons lie in the vertical plane containing the
Mo zigzag chain (sketch).
{\bf e,} Contour map of the degree of optical polarization, $\eta$({\bf k}),
as defined in the text, which depends on the energy of the photon absorbed 
in the edge transition shown in panel c. 
} 
\end{figure}

The mean-field Hamiltonian in {\bf k} space, 
$\hat{\cal{H}}(\text{\bf k})=\hat{\cal{H}}_{\text{QSH}}(\text{\bf k})
+\hat{\cal{H}}_{\text{X}}(\text{\bf k})$, 
is therefore a 4 $\times$ 4 matrix that acts on vectors 
spanned by spin-resolved $p$ and $d$
Bloch states at $\Gamma$. 
It adds the excitonic term, $\hat{\cal{H}}_{\text{X}}$, to the single particle term, $\hat{\cal{H}}_{\text{QSH}}$, based on an effective-mass model\cite{qian2014quantum} 
of GW bands: 
\begin{equation}
\hat{\cal{H}}_{\text{QSH}}=\frac{\varepsilon_p+
\varepsilon_d}{2}
\hat{\mathbb{I}}_{\tau}\otimes \hat{\mathbb{I}}_{\sigma} \,+\,
\frac{\varepsilon_p -
\varepsilon_d}{2}
\hat{\tau}_{z}\otimes \hat{\mathbb{I}}_{\sigma}\,-\,\text{Im}\{\Delta_{\text{SO}}\}
\hat{\tau}_{y}\otimes \hat{\mathbb{I}}_{\sigma}\,+\,\text{Re}\{\Delta_{\text{SO}}\}
\hat{\tau}_{x}\otimes \hat{\sigma}_x.
\label{eq:HQSH}
\end{equation}
Here the model parameters are the band inversion and effective masses
of $p$ and $d$ energy bands, $\varepsilon_p(\text{\bf k})$ and
$\varepsilon_d(\text{\bf k})$ 
[plotted in Fig.~3d as
$\xi(\text{\bf k}) = \left(\varepsilon_p -
\varepsilon_d\right)/2$ ], 
as well as 
the velocities, $v_1$ and $v_2$,
associated with the
complex spin-orbit interaction, $\Delta_{\text{SO}}(\text{\bf k})=\hbar v_2 k_y-i 
\hbar v_1 k_x$ (Figs.~3e and 3f).  
$\hat{\sigma}_{\alpha}$ and $\hat{\mathbb{I}}_{\sigma}$ are the 2 $\times$ 2
Pauli matrices and identity in spin space, whereas
$\hat{\tau}_{\alpha}$  
and $\hat{\mathbb{I}}_{\tau}$ act on the pseudospin
space of $p$ and $d$ orbital components ($\alpha=x,y,z$). The chosen parameters provide good matching between
model and first-principles GW bands (respectively dashed and dotted lines in Figs. 4a and 4b) in the {\bf k}-space region of interest. The excitonic hybridization, 
$\Delta_{\text{X}}(\text{\bf k})$, which appears in
\begin{equation}
\hat{\cal{H}}_{\text{X}} = \text{Re}\{\Delta_{\text{X}}\} 
\hat{\tau}_{x}\otimes \hat{\mathbb{I}}_{\sigma}\,-\,\text{Im}\{\Delta_{\text{X}}\}
\hat{\tau}_{y}\otimes \hat{\sigma}_x,
\label{eq:HX}
\end{equation}
is obtained numerically (Figs.~3g and 3h)
by solving two self-consistent coupled equations [see \eqref{eq:scf1} and \eqref{eq:scf2} below], ruled by the screened Coulomb interaction,
$W(\text{\bf k})$, which we extract from first principles (Supplementary Figure 1). In the QSH phase these equations
only admit the trivial solution, $\Delta_{\text{X}}=0$. The agreement
between effective-mass (Supplementary Figure 2) and first-principles (Fig.~2b) exciton wave functions points to the reliability
of the two-band model.

The conduction and
valence bands of the excitonic insulator are
respectively $E_{c\lambda}(\text{\bf k})= (\varepsilon_p+
\varepsilon_d)/2 + E_{\text{\bf k}\lambda}$ and
$E_{v\lambda}(\text{\bf k})= (\varepsilon_p+
\varepsilon_d)/2 - E_{\text{\bf k}\lambda}$, where
$E_{\text{\bf k}\lambda}=\left[ \xi^2 + \left| \Delta_{\text{SO}}\;\lambda\; \Delta_{\text{X}}\right|^2\right]^{1/2}$ and
$\lambda = \pm $ is a quantum label that reduces to the spin projection in the absence of spin-orbit
interaction. Low-lying $e-h$ excitations
close to the gap (Fig.~4c), of
energies $E_{c\lambda}(\text{\bf k})-E_{v\lambda'}(\text{\bf k})$,
occur at the pristine semimetal Fermi surface,
which is an ellipse in {\bf k} 
space obeying $\xi=0$ (white colour in Fig.~3d).
In particular, as shown in Figs.~3e-h, 
at $\Lambda$ points both $\text{Im}\{\Delta_{\text{SO}}\}$  
and $\text{Re}\{\Delta_{\text{X}}\}$ vanish, hence
$\text{Re}\{\Delta_{\text{SO}}\}$ 
and $\text{Im}\{\Delta_{\text{X}}\}$ add quadratically
without interfering, resulting in the
approximate gap value
$E_{c\lambda}(\Lambda)-E_{v\lambda'}(\Lambda)
=2\left[(\text{Re}\{\Delta_{\text{SO}}\})^2
+ ( \text{Im}\{\Delta_{\text{X}}\} )^2 \right]^{1/2}$. 
Therefore, the insulator is
simultaneously excitonic and 
topological (QSHX), 
as an effect of the self-organization of $\Delta_{\text{X}}$({\bf k}). This may be checked most easily by adiabatic continuation,
since the pristine QSH gap smoothly increases as the temperature is lowered below the critical value of the QSHX phase, around 700 K 
(blue curve in Fig.~5b). We explicitly calculate $\mathbb{Z}_2$ in 
Supplementary Note 1.

\begin{figure}[htbp]
\setlength{\unitlength}{1 cm}
\includegraphics[trim=0cm 0cm 0 0,clip=true,width=16.0cm]{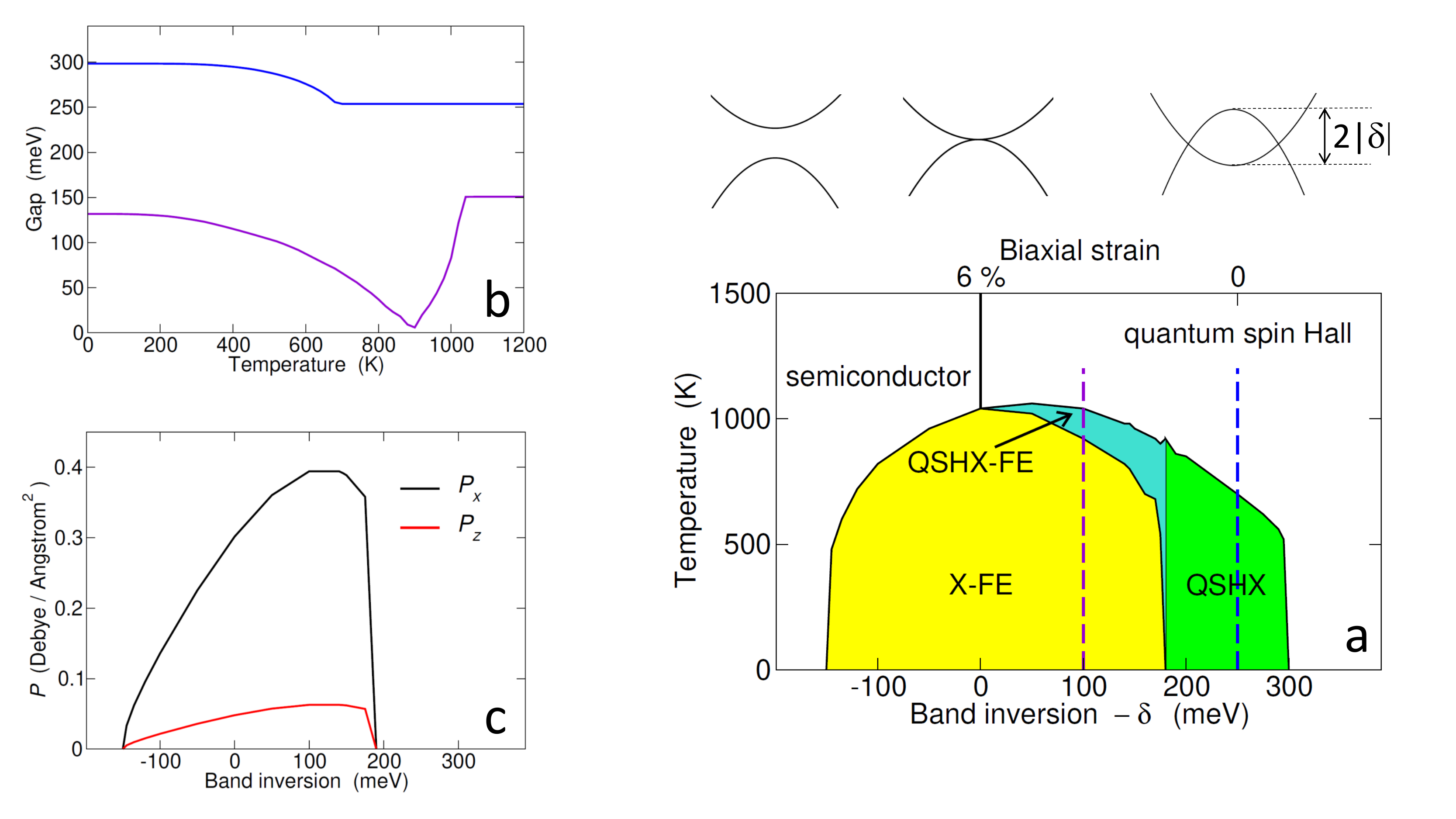}
\caption{
{\bf Phase diagram vs strain and temperature.} 
{\bf a,} Phase diagram in the temperature -- 
biaxial strain space. Strain is mapped into
band inversion, $-\delta$, 
as illustrated by the insets (in the sketches $\Delta=0$). 
QSHX labels the topological excitonic insulator 
(predicted at zero strain), QSHX-FE the ferroelectric 
topological excitonic phase, and
X-FE the topologically trivial, ferroelectric excitonic phase.
{\bf b,} Energy gap vs temperature along selected cuts in the phase
diagram (violet and blue dashed lines in panel a), corresponding respectively to 
$-\delta=100$ meV (violet curve) and $-\delta=250$ meV (blue curve). 
The QSHX (-FE) gap is larger (smaller)
than the QSH gap at high temperature due to the lack of interference 
(destructive interference) between excitonic and topological orders.   
{\bf c,} Permanent electric dipole along $x$ and $z$ direction vs band 
inversion at zero temperature.
}
\end{figure}

\noindent {\bf Fingerprints} The excitonic hybridization 
spontaneously breaks the inversion symmetry of the crystal ($\hat{\cal{H}}_{\text{X}}$ anticommutes with the inversion operator,
$\hat{I}=-\hat{\tau}_z\otimes\hat{\mathbb{I}}_{\sigma}$), which leads to clear-cut observable properties. 
The Kramers degeneracy of bands is lifted, as shown in Fig.~4b
(black and red solid lines point to the two spin projections),
with band splittings reaching a maximum of about 20 meV for $k_x \sim 0.05 - 0.1$ [units of $(2\pi)/a_x$] and vanishing on the $\Gamma\Lambda$ line 
(solid black line in Fig.~4a), as a consequence of the two-fold rotational symmetry that restores degeneracy. The QSHX phase exhibits circular dichroism, like  
monolayer $H$-MoS$_2$ (Refs.~\citeonline{Cao2012,Urbaszek2018}): the optical absorption of 
a photon whose 
wave vector lies in the $yz$ plane (Fig.~4d), $\cal{A}$, 
depends on the
circular polarization of the photon itself, $\sigma^+$ or $\sigma^-$.
Figure 4e shows the {\bf k}-resolved degree of optical polarization\cite{Cao2012},
\begin{equation}
    \eta(\text{\bf k}) =\frac{{\cal{A}}(\sigma^+)-{\cal{A}}(\sigma^-)}{{\cal{A}}(\sigma^+)+{\cal{A}}(\sigma^-)},
    \label{eq:eta}
\end{equation}
evaluated for the optical transition from the highest valence to the lowest conduction band, whose excitation energy,   
$E_{c\lambda}(\text{\bf k})-E_{v\lambda}(\text{\bf k})$, 
is shown in Fig.~4c.
Remarkably, as $\eta$({\bf k}) circles around the semimetal Fermi surface, it modulates from its
lower to its upper bound, respectively $-1$ (blue color in Fig.~4e) and 
$+1$ (red colour). These limit values are effectively optical selection rules coupling orbital and spin degrees of freedom, like in the case  
of monolayer $H$-MoS$_2$. On the contrary,
in the QSH ground state $\eta$({\bf k}) $=0$ at each {\bf k} point, 
since the microscopic transverse currents, which are responsible of the net angular momentum, exactly cancel out due to Kramers degeneracy\cite{Yafet1963}. We expect the dichroic signal to survive to final-state interactions, not included in our calculation (Methods), and hence disclose the intrinsic chirality of the QSHX ground state.

\noindent {\bf Excitonic phases and ferroelectricity} The application of
biaxial strain to $T'$-MoS$_2$ is a practical handle to tune the
band inversion\cite{qian2014quantum} and hence the energy scales 
ruling the ground state, as shown in Fig.~5a. 
Excitonic correlations tend to destroy topological order
as the band overlap is suppressed, since $\Delta_{\text{SO}}$ 
decreases while
$\Delta_{\text{X}}$ increases. Such balance 
eventually leads to 
a topologically trivial excitonic insulator (region 
in yellow colour, X-FE), after crossing a
phase allowing for destructive interference
between $\Delta_{\text{SO}}$ and $\Delta_{\text{X}}$
(cyan colour, QSHX-FE). This tiny intermediate region,
located around the value of 180 meV, is broadened by temperature
up to the frontier with the QSH phase.  
The topological character of each phase
is made evident by the cuts of the phase diagram along the temperature
axis (violet and blue dashed lines in Fig.~5a) displayed in Fig.~5b.
Starting from the QSH phase and lowering the temperature, the gap increases
when entering the QSHX phase (blue curve) but decreases while
crossing the QSHX-FE region (violet curve), until the gap closes and opens again in the, now topologically trivial, X-FE phase 
(cf.~Supplementary Note 1).

Both QSHX-FE and X-FE phases totally distort the pristine
$C_{2h}$ symmetry by breaking the screw axis 
symmetry along $y$ (Supplementary Figure 4), in addition to inversion. 
This is related to the macroscopic condensation of 
the exciton electric dipole in the $xz$ plane, which makes the system ferroelectric\cite{Portengen1996b}. 
Contrary to usual ferroelectrics, like BaTiO$_3$, here the permanent electric dipole, {\bf P}, is not due to the displacement of anions and
cations but to the modulation of the electronic charge
density associated with the exciton polarization (Fig.~5c). 
This may open fascinating new routes, like the realization
of ultrafast 
switches between conductive and insulating (ferroelectric) 
beahvior, locally controllable by strain or screening, 
or the exploration of exotic electronic collective modes\cite{Portengen1996b}, coherently radiating in the THz range.

In conclusion, we have demonstrated that a paradigmatic   
two-dimensional topological insulator
is also an excitonic insulator,
by combining calculations from first principles with a self-consistent
mean-field model. We expect our results to be relevant to other $T'$ polytypes, 
including those that are semimetal according to MBPT, like WTe$_2$, 
which are harder to simulate in view of the enhanced screening of $e-h$ interactions.    
In particular, WTe$_2$ owns the record temperature of 100 K for the QSH effect\cite{Jarillo-Herrero2018} and exhibits unexpected but intriguing properties that might be related
to the excitonic insulator, like ferroelectricity\cite{fei-cobden_2018}
and gate-induced superconductivity\cite{Fatemi2018,Sajadi2018} in close proximity to the insulating phase.

\begin{methods}

\subsection{Computational details of many-body
perturbation theory from first principles.}

\emph{The ground-state structure and Kohn-Sham states} were calculated for a single layer of $T'$-MoS$_2$ by using a DFT approach, as implemented in the Quantum ESPRESSO package\cite{Giannozzi2009}.
The generalized gradient approximation (GGA) PBE
parametrization\cite{perdew1996generalized}
was adopted together with plane wave basis set and norm-conserving
pseudopotentials to model the electron-ion interaction. 
Fully relativistic pseudopotentials treating the $sp$ semicore states of the transition metal atoms as valence electrons were employed.
The kinetic
energy cutoff for the wave functions was set to 90 Ry.
The Brillouin zone was sampled by using a  16 $\times$ 16 $\times$
1 {\bf k}-point grid. 
The supercell size perpendicular to the $T'$-MoS$_2$ layer was set to $a_z=15.98$ {\AA} and checked to be large enough to avoid spurious interactions
with its replica.

\noindent \emph{The equilibrium atomic lattice parameters} were obtained by performing a full relaxation of the cell and atomic positions. 
The obtained equilibrium lattice parameters, $a_x= 5.74$ \AA, $a_y= 3.19$ \AA, as well as the Kohn-Sham electronic gap were in very good agreement with previous literature\cite{qian2014quantum}.

\noindent \emph{Many-body perturbation theory}\cite{Onida-Reining-Rubio_2002} calculations
were performed using the Yambo code\cite{Marini2009,sangalli2019}.
Many-body corrections to the Kohn-Sham eigenvalues were
calculated within the $G0W0$ approximation to the self-energy operator,
where the dynamic dielectric function was obtained within the
plasmon-pole approximation\cite{Godby1989}.
The spectrum of excited states was then computed by solving the
Bethe-Salpeter equation (BSE). The static screening in the direct
term was calculated within the random-phase approximation with
inclusion of local field effects; the Tamm-Dancoff approximation
for the BSE Hamiltonian was employed after having verified that
the correction introduced by coupling the resonant and antiresonant
part was negligible.
Converged excitation energies were obtained
considering respectively 2 valence and 2 conduction bands in the
BSE matrix.
For the calculations of the GW band structure and the
Bethe-Salpeter matrix, the Brillouin zone was sampled with a
70 $\times$ 35 $\times$ 1 {\bf k}-point grid.
A kinetic energy cutoff of 60 Ry was used for the evaluation of
the exchange part of the self energy and 10 Ry for the screening
matrix size. 248 unoccupied bands were used to build the polarizability and integrate the self-energy.
The Coulomb interaction was truncated\cite{Rozzi2006} in the layer-normal direction to avoid spurious interactions with the image systems. Note that the GW gap obtained in this work (0.26 eV) is similar to the value reported in Ref.~\citeonline{kan2014structures} calculated at HSE06 level (0.23 eV). 
A smaller value of the $G0W0$ gap (0.08 eV) is instead reported in Ref.~\citeonline{qian2014quantum}. The observed discrepancy can be probably explained by the use of a non-truncated Coulomb potential in that work.
The approach used in the GW and BSE calculations takes into account the full spinorial nature of the electronic Kohn-Sham states, providing superior accuracy than perturbative treatments of spin-orbit coupling.

\noindent {\bf Two-band model and self-consistent mean-field theory.} 
The effective-mass Hamiltonian, 
$\hat{\cal{H}}(\text{\bf k})=\hat{\cal{H}}_{\text{QSH}}(\text{\bf k})
+\hat{\cal{H}}_{\text{X}}(\text{\bf k})$, 
acts on four-component vectors, $\left(u_{\uparrow}(\text{\bf k}),v_{\uparrow}(\text{\bf k}),u_{\downarrow}(\text{\bf k}),v_{\downarrow}(\text{\bf k})\right)$,
with $u_{\sigma}$({\bf k}) and $v_{\sigma}$({\bf k}) 
being the spin- and {\bf k}-resolved envelope functions\cite{LuttingerKohn1955} of 
the $p_y$ and $d_{yz}$
Bloch states at $\Gamma$, respectively, and 
$\sigma = \uparrow,\downarrow$. 
The vector normalization is such that
$\sum_{\sigma}(\left|u_{\sigma}(\text{\bf k})\right|^2 + \left|v_{\sigma}(\text{\bf k})\right|^2) = 1$.

\noindent \emph{QSH Hamiltonian.} The non-interacting term of Eq.~\eqref{eq:HQSH},  $\hat{\cal{H}}_{\text{QSH}}$, is modelled after Ref.~2
to provide the bands of
the QSH insulator---renormalized within the GW approximation---by 
accounting for the
spin-orbit interaction. It complies with the $C_{2h}$ point symmetry group,
which includes the inversion,
$\hat{I}=-\hat{\tau}_z\otimes\hat{\mathbb{I}}_{\sigma}$,
and the two-fold screw axis rotation along $y$,
$\hat{C}_{2y}=i\,\hat{\tau}_z\otimes\hat{\sigma}_{y}$.
The band dispersions  
are $\varepsilon_p(\text{\bf k})=
-\delta \; - \hbar^2 k_x^2 / (2m_{px}) \;- \hbar^2 k_y^2 / (2m_{py})  $
and
$\varepsilon_d(\text{\bf k})=
\delta\;  + \hbar^2 k_x^2 / (2m_{dx}) \;+ \hbar^2 k_y^2 / (2m_{dy})  $, 
with $-\delta$ being the band inversion.
We optimize the matching with first-principles GW bands by choosing
$\delta = -250$ meV, $m_{px}/m_e=0.6$, $m_{py}/m_e=0.31$,
$m_{dx}/m_e=2.48$, $m_{dy}/m_e=0.5$,
$v_1 =$ 1.5$\cdot 10^{15}$ \AA  /s, $v_2 =$ 1.3$\cdot 10^{15}$ \AA  /s
($m_e$ is the free electron mass).

\noindent \emph{Excitonic hybridization.} The mean-field excitonic Hamiltonian, $\hat{\cal{H}}_{\text{X}}$({\bf k}), 
in principle
accounts for eight independent order parameters, i.e., the four complex hybridization terms coupling all possible pairs
of conduction- and valence-band spin projections.
The form \eqref{eq:HX} complies with
the constrain that $\hat{\cal{H}}_{\text{X}}$ 
is even in 
{\bf k}, $\hat{\cal{H}}_{\text{X}}$({\bf k})
$= \hat{\cal{H}}_{\text{X}}(-\text{\bf k})$,
and invariant under time reversal, 
$\hat{\Theta} \hat{\cal{H}}_{\text{X}}(-\text{\bf k}) = 
\hat{\cal{H}}_{\text{X}}(\text{\bf k}) \hat{\Theta}$, 
where
$\hat{\Theta} = i\, \hat{\mathbb{I}}_{\tau}\otimes \hat{\sigma}_y \hat{K}$ and $\hat{K}$
are the time-reversal
and the complex conjugation operator, respectively.
The excitonic hybridization terms are derived from
the interband Coulomb interaction
through the usual mean-field decoupling
procedure, as
\begin{equation}
   \text{Re}\{\Delta_{\text{X}}(\text{\bf k})\}=-\sum_{\text{\bf k}'}
   W(\text{\bf{k}}-\text{\bf k}')\left< 
   p^{\dagger}_{\text{\bf k}'\uparrow}  
   d_{\text{\bf k}'\uparrow}
   \right>,
   \label{eq:DeltaRe}
\end{equation}
and
\begin{equation}
   \text{Im}\{\Delta_{\text{X}}(\text{\bf k})\}=-\sum_{\text{\bf k}'}
   W(\text{\bf{k}}-\text{\bf k}')\left< 
   p^{\dagger}_{\text{\bf k}'\uparrow}  
   d_{\text{\bf k}'\downarrow}
   \right>,
   \label{eq:DeltaIm}
\end{equation}
where the Fermi field operators $p_{\text{\bf k}\sigma}$ 
and $d_{\text{\bf k}\sigma}$
destroy electrons occupying $p$ and $d$ Bloch states, respectively, and the symbol $\left<\ldots\right>$ stands for the
quantum statistical average 
[the expressions \eqref{eq:DeltaRe} and \eqref{eq:DeltaIm} do not change
as one reverses the spins of
$p^{\dagger}_{\text{\bf k}\sigma}$ and $d_{\text{\bf k}\sigma'}$ simultaneously].
Whereas quantum and thermal fluctuations are believed
to disrupt long range order in low dimensions, here we expect that
excitonic correlations are stabilized by the long range of the Coulomb interaction, as verified in the one-dimensional case through 
extensive quantum Monte Carlo simulations\cite{Varsano-Rontani2017}. 

\noindent \emph{Dressed interaction.} The key quantity above is the {\bf q}-resolved screened Coulomb
interaction\cite{Cudazzo2011}, $W(\text{\bf q}) =
V_0(\text{\bf q})/(1 + 2\pi \alpha_{\text{2D}}
\left| \text{\bf q}\right|)$, which we fit to the one
obtained from first principles within the random phase approximation
(Supplementary Figure 1). The long-wavelength term, 
$ V_0(\text{\bf q}) \sim 1/\left| \text{\bf q}\right|$, is unscreened in
two dimensions\cite{Cudazzo2011} and hence  
the dominant contribution 
to $e-h$ attraction. Since the prefactor of $1/\left| \text{\bf q}\right|$ 
is determined by the 
dimensions of the supercell used in the first-principles 
calculation\cite{Rozzi2006}, i.e.,
$a_x=5.743$ \AA, $a_y=3.191$ \AA, $a_z=15.98$ \AA, the only  
free parameter   
is the two-dimensional polarizability\cite{Cudazzo2011}, $\alpha_{\text{2D}}$. We adjust $\alpha_{\text{2D}}$ 
to match
effective-mass and first-principles exciton binding energies,
with $\alpha_{\text{2D}}=10.75$ 
corresponding to the three-dimensional dielectric constant of 9.5 for the stack of $T'$-MoS$_2$ monolayers---the actual bulk of the supercell calculation\cite{Cudazzo2011}. This latter figure
reasonably compares with the values---between 10 and 20---assumed by the
first-principles dielectric function in
the $q$-range of $0.02\div 0.05$ [units of $(2 \pi)/a_x$].

\noindent \emph{Eigenvectors.} The eigenvectors of $\hat{\cal{H}}(\text{\bf k})$ are the conduction and valence band states of the correlated insulator.
The first of the two valence bands, $(v,\lambda=+)$, has as
wave function envelopes 
$u_{\sigma}(\text{\bf k},v,+)= u_0/\sqrt{2}$
and $v_{\sigma}(\text{\bf k},v,+)= iv_0\exp(-i\varphi)/\sqrt{2}$, with
$u_0$ and $v_0$ positive numbers whose magnitudes
are given by $u_0^2=1/2[1-\xi(\text{\bf k})/E_{\text{\bf k}+}] $ and $u_0^2 + v_0^2 =1$;
the phase is provided by
$iu_0v_0 \exp(i\varphi) = (\Delta_{\text{SO}} +
\Delta_{\text{X}})/2E_{\text{\bf k}+} $.
The second band, $(v,\lambda = -)$,
has
$u_{\sigma}(\text{\bf k},v,-)= -\sigma u_0'/\sqrt{2}$ and $v_{\sigma}(\text{\bf k},v,-)= -\sigma iv_0'\exp(i\varphi')/\sqrt{2}$ 
(the spin index takes the values $\sigma = \pm$ when occurring in the body of a formula),
with $u_0'^2=1/2[1-\xi(\text{\bf k})/E_{\text{\bf k}-}] $, 
$u_0'^2 + v_0'^2 =1$, and
$iu_0'v_0' \exp(i\varphi') = (\Delta_{\text{SO}} 
-\Delta_{\text{X}})/2E_{\text{\bf k}-}$.
The envelopes of conduction bands
$(c,\lambda)$ have similar 
expressions, which are obtained from
$u_{\sigma}(\text{\bf k},v,\lambda)$ and $v_{\sigma}(\text{\bf k},v,\lambda)$
by swapping $u_0$ for $v_0$ 
($u_0'$ for $v_0'$) in the
formulae and
simultaneously adding $\pi$ to $\varphi$ ($\varphi'$).

\noindent \emph{Self-consistent equations.} The many-body ground state, $\left|\Psi_0\right>$, 
is the Slater determinant with all valence band states filled, i.e., $\left|\Psi_0\right> = \prod_{\text{\bf k}}
\gamma^{\dagger}_{\text{\bf k},v,+}\gamma^{\dagger}_{\text{\bf k},v,-}\left|0\right>$, where $\left|0\right>$
is the vacuum and $\gamma_{\text{\bf k},\alpha,\lambda} = \sum_{\sigma} 
u_{\sigma}(\text{\bf k},\alpha,\lambda) \,p_{\text{\bf k}\sigma}
+ v_{\sigma}(\text{\bf k},\alpha,\lambda) \,d_{\text{\bf k}\sigma}$
is the Fermi operator destroying an electron occupying the eigenstate of 
$\hat{\cal{H}}(\text{\bf k})$ of wave vector {\bf k}
and band index $(\alpha,\lambda)$, with $\alpha=c,v$.
The knowledge of many-body states, together with 
Eqs.~\eqref{eq:DeltaRe} and \eqref{eq:DeltaIm}, 
allows us to write explicitly the self-consistent equations for
$\Delta_{\text{X}}$,
\begin{eqnarray}
     \text{Re}\{\Delta_{\text{X}}(\text{\bf k})\} 
       &=& \frac{1}{4} \sum_{\text{\bf k}'} W(\text{\bf{k}}-\text{\bf k}') 
     \bigg[
     \frac{ \text{Re}\{\Delta_{\text{X}}(\text{\bf k}')\} + \hbar v_2 k_y'  }{ E_{\text{\bf k}'+}   }
     \left[f_F\!\left(E_{v+}(\text{\bf k}')\right) - f_F\left(E_{c+}(\text{\bf k}')\right) \right] \nonumber \\ 
     &+&
     \frac{ \text{Re}\{\Delta_{\text{X}}(\text{\bf k}')\} - \hbar v_2 k_y'  }{ E_{\text{\bf k}'-}   }
     \left[f_F\!\left(E_{v-}(\text{\bf k}')\right) - f_F\!\left(E_{c-}(\text{\bf k}')\right) \right]  \bigg] ,
     \label{eq:scf1}
\end{eqnarray}
and
\begin{eqnarray}
     \text{Im}\{\Delta_{\text{X}}(\text{\bf k})\} 
       &=& \frac{1}{4} \sum_{\text{\bf k}'} W(\text{\bf{k}}-\text{\bf k}') 
     \bigg[
     \frac{ \text{Im}\{\Delta_{\text{X}}(\text{\bf k}')\} - \hbar v_1 k_x'  }{ E_{\text{\bf k}'+}   }
     \left[f_F\!\left(E_{v+}(\text{\bf k}')\right) - f_F\left(E_{c+}(\text{\bf k}')\right) \right] \nonumber \\ 
     &+&
     \frac{ \text{Im}\{\Delta_{\text{X}}(\text{\bf k}')\} + \hbar v_1 k_x'  }{ E_{\text{\bf k}'-}   }
     \left[f_F\!\left(E_{v-}(\text{\bf k}')\right) - f_F\!\left(E_{c-}(\text{\bf k}')\right) \right]  \bigg].
     \label{eq:scf2}
\end{eqnarray}
Here $f_F(E)=(\exp{\beta(E-\mu)} + 1)^{-1}$
is the Fermi distribution function, $\beta = 1/k_BT$ is inversely proportional to the temperature $T$, $k_B$ is Boltzmann's constant, and $\mu$ is the equilibrium chemical potential, which we place at midgap neglecting small deviations expected at finite temperature.
Note that the divergence of $W$ at long wavelength is
harmless since the interaction is integrated
over the Brillouin zone.
Equations \eqref{eq:scf1} and \eqref{eq:scf2}, which
are coupled together essentially through the denominators $E_{\text{\bf k}'\lambda}$, allow for the trivial 
solution 
$\Delta_{\text{X}}(\text{\bf k})\equiv 0$ in the QSH phase. The non trivial solution is obtained 
numerically through recursion, using the {\bf k}-resolved exciton wave function as a first-iteration seed\cite{Varsano-Rontani2017}, which allows for quick and robust convergence. 

\noindent \emph{Topological invariant $\mathbb{Z}_2$.} We derive the topological
invariant $\mathbb{Z}_2$ by applying the test originally developed by Kane and Mele for
graphene\cite{Kane2005b} to $\hat{\cal{H}}(\text{\bf k})$. 
This relies on the evaluation of
the overlap, $P(\text{\bf k})$, between
the single-particle state $(\text{\bf k},v,\lambda)$ and the time reversed of $(\text{\bf k},v,-\lambda)$,
\begin{equation}
    P(\text{\bf k}) = \left< \text{\bf k},v,\lambda \right| \hat{\Theta} \left| 
    \text{\bf k},v,-\lambda \right>_{\text{sp}}
    = u_0u_0'-v_0v_0'e^{i(\varphi-\varphi')},
    \label{eq:Z2}
\end{equation}
with $\left<...\right>_{\text{sp}}$ being the scalar product between four-component vectors.
In Supplementary Note 1 we assess the number of pairs of zeroes of $P(\text{\bf k})$ occurring 
at ${\text{\bf k}}^*$ and $-{\text{\bf k}}^*$, which at once characterizes the
topology of the many-body state.

\noindent \emph{Circular dichroism and permanent electric dipole.} The degree of optical polarization, $\eta$({\bf k}), is defined by Eq.~\eqref{eq:eta} through
the optical absorption, ${\cal A}(\sigma^{\pm})$,  of a photon having circular polarization, $\sigma^+$ or $\sigma^-$, and
wave vector lying in the $yz$ plane. We evaluate ${\cal A}(\sigma^{\pm})$
through Fermi golden rule,
\begin{equation}
  {\cal A}(\sigma^{\pm}) = \frac{2\pi}{\hbar}
  \sum_{\lambda} \left|\left<\Psi_0\right|
  {\cal{D}}(\sigma^{\pm})\gamma^{\dagger}_{\text{\bf k}c\lambda}
  \gamma_{\text{\bf k}v\lambda}
  \left|\Psi_0\right>\right|^2\delta\!\left(\hbar\omega - E_{c}(\text{\bf k}) + E_{v}(\text{\bf k}) \right),
  \label{eq:Fermi_golden}
\end{equation}
by selecting the  
mimimum $e-h$ pair excitation energy available 
at a given point in {\bf k} space, $E_{c}(\text{\bf k}) - E_{v}(\text{\bf k})$, i.e., 
the photon energy
$\hbar \omega$ depends on {\bf k}, as shown in Fig.~4c. The light-matter
interaction within the dipole approximation,
\begin{equation}
{\cal{D}}(\sigma^{\pm}) = 
e\left({\cal{E}}_x x_{\Gamma} \pm i 
{\cal{E}}_z z_{\Gamma} \right)
\sum_{\text{\bf k}\sigma} \left(
p^{\dagger}_{\text{\bf k}\sigma}d_{\text{\bf k}\sigma} + d_{\text{\bf k}\sigma}^{\dagger}p_{\text{\bf k}\sigma}
\right), 
\end{equation}
depends on the (real) 
coordinate interband matrix elements evaluated at $\Gamma$, 
$x_{\Gamma}$ and $z_{\Gamma}$, which we
extract from first principles ($x_{\Gamma}=1.94$ \AA, $z_{\Gamma}=0.308$
\AA, $y_{\Gamma}=0$ due to symmetry). Here ${\cal{E}}_x={\cal{E}}_z={\cal{E}}$ is the electric field, and the sign $\pm$ picks up the helicity of the field that is circularly polarized in the $xz$ plane. 
The permanent electric dipole, $\left<\text{\bf P}\right>$,
is the equilibrium average of the periodic part of the interband dipole operator over the many-body
ground state\cite{Portengen1996b},
\begin{equation}
  \left<\text{\bf P}\right> = 
  \frac{ e\left(x_{\Gamma}\text{\bf i} + z_{\Gamma}\text{\bf k}\right) }{ N_xa_xN_ya_y }
    \sum_{\text{\bf k}\sigma}\left<  p^{\dagger}_{\text{\bf k}\sigma}d_{\text{\bf k}\sigma} + d_{\text{\bf k}\sigma}^{\dagger}
    p_{\text{\bf k}\sigma} \right>,
\end{equation}
where $N_xN_y$ is the number of unit cells used in the simulation.

\end{methods}

\section*{References}


\begin{thebibliography}{10}
\expandafter\ifx\csname url\endcsname\relax
  \def\url#1{\texttt{#1}}\fi
\expandafter\ifx\csname urlprefix\endcsname\relax\def\urlprefix{URL }\fi
\providecommand{\bibinfo}[2]{#2}
\providecommand{\eprint}[2][]{\url{#2}}

\bibitem{Kane2005a}
\bibinfo{author}{Kane, C.~L.} \& \bibinfo{author}{Mele, E.~J.}
\newblock \bibinfo{title}{Quantum spin {H}all effect in graphene}.
\newblock \emph{\bibinfo{journal}{Phys. Rev. Lett.}}
  \textbf{\bibinfo{volume}{95}}, \bibinfo{pages}{226801}
  (\bibinfo{year}{2005}).
\newblock
  \urlprefix\url{https://link.aps.org/doi/10.1103/PhysRevLett.95.226801}.

\bibitem{qian2014quantum}
\bibinfo{author}{Qian, X.}, \bibinfo{author}{Liu, J.}, \bibinfo{author}{Fu, L.}
  \& \bibinfo{author}{Li, J.}
\newblock \bibinfo{title}{Quantum spin {H}all effect in two-dimensional
  transition metal dichalcogenides}.
\newblock \emph{\bibinfo{journal}{Science}} \textbf{\bibinfo{volume}{346}},
  \bibinfo{pages}{1344--1347} (\bibinfo{year}{2014}).

\bibitem{Jarillo-Herrero2018}
\bibinfo{author}{Wu, S.} \emph{et~al.}
\newblock \bibinfo{title}{Observation of the quantum spin {H}all effect up to
  100 {K}elvin in a monolayer crystal}.
\newblock \emph{\bibinfo{journal}{Science}} \textbf{\bibinfo{volume}{359}},
  \bibinfo{pages}{76--79} (\bibinfo{year}{2018}).
\newblock \urlprefix\url{http://science.sciencemag.org/content/359/6371/76}.
\newblock \eprint{http://science.sciencemag.org/content/359/6371/76.full.pdf}.

\bibitem{Onida-Reining-Rubio_2002}
\bibinfo{author}{Onida, G.}, \bibinfo{author}{Reining, L.} \&
  \bibinfo{author}{Rubio, A.}
\newblock \bibinfo{title}{{Electronic excitations: density-functional versus
  many-body Green's-function approaches}}.
\newblock \emph{\bibinfo{journal}{Rev. Mod. Phys.}}
  \textbf{\bibinfo{volume}{74}}, \bibinfo{pages}{601--659}
  (\bibinfo{year}{2002}).

\bibitem{Keldysh1964}
\bibinfo{author}{Keldysh, L.~V.} \& \bibinfo{author}{Kopaev, Y.~V.}
\newblock \bibinfo{title}{Possible instability of the semimetallic state
  against {C}oulomb interaction}.
\newblock \emph{\bibinfo{journal}{Fiz. Tverd. Tela}}
  \textbf{\bibinfo{volume}{6}}, \bibinfo{pages}{2791} (\bibinfo{year}{1964}).
\newblock \bibinfo{note}{[Sov. Phys. Sol. State {\bf 6,} 2219 (1965)]}.

\bibitem{Cloizeaux1965}
\bibinfo{author}{des Cloizeaux, J.}
\newblock \bibinfo{title}{Excitonic instability and crystallographic anomalies
  in semiconductors}.
\newblock \emph{\bibinfo{journal}{J. Phys. Chem. Solids}}
  \textbf{\bibinfo{volume}{26}}, \bibinfo{pages}{259} (\bibinfo{year}{1965}).

\bibitem{Kohn1967}
\bibinfo{author}{J{\`e}rome, D.}, \bibinfo{author}{Rice, T.~M.} \&
  \bibinfo{author}{Kohn, W.}
\newblock \bibinfo{title}{Excitonic insulator}.
\newblock \emph{\bibinfo{journal}{Phys. Rev.}} \textbf{\bibinfo{volume}{158}},
  \bibinfo{pages}{462} (\bibinfo{year}{1967}).

\bibitem{Halperin1968}
\bibinfo{author}{Halperin, B.~I.} \& \bibinfo{author}{Rice, T.~M.}
\newblock \bibinfo{title}{The excitonic state at the semiconductor-semimetal
  transition}.
\newblock \emph{\bibinfo{journal}{Solid State Phys.}}
  \textbf{\bibinfo{volume}{21}}, \bibinfo{pages}{115} (\bibinfo{year}{1968}).

\bibitem{Volkov1973}
\bibinfo{author}{Volkov, V.~A.} \& \bibinfo{author}{Kopaev, Y.~V.}
\newblock \bibinfo{title}{Theory of phase transitions in semiconductors of the
  {A}$_4${B}$_6$ group}.
\newblock \emph{\bibinfo{journal}{Zh. Eksp. i Teor. Fiz.}}
  \textbf{\bibinfo{volume}{64}}, \bibinfo{pages}{2184--2915}
  (\bibinfo{year}{1973}).
\newblock \bibinfo{note}{[Sov. Phys.--JETP {\bf 37,} 1103-1108 (1974)]}.

\bibitem{Bernevig2006}
\bibinfo{author}{Bernevig, B.~A.}, \bibinfo{author}{Hughes, T.~L.} \&
  \bibinfo{author}{Zhang, S.-C.}
\newblock \bibinfo{title}{Quantum spin {H}all effect and topological phase
  transition in {H}g{T}e quantum wells}.
\newblock \emph{\bibinfo{journal}{Science}} \textbf{\bibinfo{volume}{314}},
  \bibinfo{pages}{1757--1761} (\bibinfo{year}{2006}).
\newblock \urlprefix\url{http://science.sciencemag.org/content/314/5806/1757}.
\newblock
  \eprint{http://science.sciencemag.org/content/314/5806/1757.full.pdf}.

\bibitem{Konig2007}
\bibinfo{author}{K{\"o}nig, M.} \emph{et~al.}
\newblock \bibinfo{title}{Quantum spin {H}all insulator state in {H}g{T}e
  quantum wells}.
\newblock \emph{\bibinfo{journal}{Science}} \textbf{\bibinfo{volume}{318}},
  \bibinfo{pages}{766--770} (\bibinfo{year}{2007}).
\newblock \urlprefix\url{http://science.sciencemag.org/content/318/5851/766}.
\newblock \eprint{http://science.sciencemag.org/content/318/5851/766.full.pdf}.

\bibitem{Liu2008}
\bibinfo{author}{Liu, C.}, \bibinfo{author}{Hughes, T.~L.},
  \bibinfo{author}{Qi, X.-L.}, \bibinfo{author}{Wang, K.} \&
  \bibinfo{author}{Zhang, S.-C.}
\newblock \bibinfo{title}{Quantum spin {H}all effect in inverted type-{II}
  semiconductors}.
\newblock \emph{\bibinfo{journal}{Phys. Rev. Lett.}}
  \textbf{\bibinfo{volume}{100}}, \bibinfo{pages}{236601}
  (\bibinfo{year}{2008}).

\bibitem{Knez2011}
\bibinfo{author}{Knez, I.}, \bibinfo{author}{Du, R.-R.} \&
  \bibinfo{author}{Sullivan, G.}
\newblock \bibinfo{title}{Evidence for helical edge modes in inverted
  {I}n{A}s/{G}a{S}b quantum wells}.
\newblock \emph{\bibinfo{journal}{Phys. Rev. Lett.}}
  \textbf{\bibinfo{volume}{107}}, \bibinfo{pages}{136603}
  (\bibinfo{year}{2011}).

\bibitem{Budich2014}
\bibinfo{author}{Budich, J.~C.}, \bibinfo{author}{Trauzettel, B.} \&
  \bibinfo{author}{Michetti, P.}
\newblock \bibinfo{title}{Time reversal symmetric topological exciton
  condensate in bilayer {H}g{T}e quantum wells}.
\newblock \emph{\bibinfo{journal}{Phys. Rev. Lett.}}
  \textbf{\bibinfo{volume}{112}}, \bibinfo{pages}{146405}
  (\bibinfo{year}{2014}).
\newblock
  \urlprefix\url{https://link.aps.org/doi/10.1103/PhysRevLett.112.146405}.

\bibitem{Pikulin2014}
\bibinfo{author}{Pikulin, D.~I.} \& \bibinfo{author}{Hyart, T.}
\newblock \bibinfo{title}{Interplay of exciton condensation and the quantum
  spin {H}all effect in $\mathrm{InAs}/\mathrm{GaSb}$ bilayers}.
\newblock \emph{\bibinfo{journal}{Phys. Rev. Lett.}}
  \textbf{\bibinfo{volume}{112}}, \bibinfo{pages}{176403}
  (\bibinfo{year}{2014}).
\newblock
  \urlprefix\url{https://link.aps.org/doi/10.1103/PhysRevLett.112.176403}.

\bibitem{Du2017}
\bibinfo{author}{Du, L.} \emph{et~al.}
\newblock \bibinfo{title}{{Evidence for a topological excitonic insulator in
  {I}n{A}s/{G}a{S}b bilayers}}.
\newblock \emph{\bibinfo{journal}{{Nature Commun.}}}
  \textbf{\bibinfo{volume}{{8}}}, \bibinfo{pages}{{1971}}
  (\bibinfo{year}{{2017}}).

\bibitem{Xue2018}
\bibinfo{author}{Xue, F.} \& \bibinfo{author}{MacDonald, A.~H.}
\newblock \bibinfo{title}{{Time-Reversal Symmetry-Breaking Nematic Insulators
  near Quantum Spin {H}all Phase Transitions}}.
\newblock \emph{\bibinfo{journal}{Phys. Rev. Lett.}}
  \textbf{\bibinfo{volume}{{120}}}, \bibinfo{pages}{186802}
  (\bibinfo{year}{{2018}}).

\bibitem{Zhu2019}
\bibinfo{author}{Zhu, Q.}, \bibinfo{author}{Tu, M. W.-Y.},
  \bibinfo{author}{Tong, Q.} \& \bibinfo{author}{Yao, W.}
\newblock \bibinfo{title}{Gate tuning from exciton superfluid to quantum
  anomalous {H}all in van der {W}aals heterobilayer}.
\newblock \emph{\bibinfo{journal}{Science Advances}}
  \textbf{\bibinfo{volume}{5}}, \bibinfo{pages}{eaau6120}
  (\bibinfo{year}{2019}).
\newblock \urlprefix\url{http://advances.sciencemag.org/content/5/1/eaau6120}.
\newblock
  \eprint{http://advances.sciencemag.org/content/5/1/eaau6120.full.pdf}.

\bibitem{Portengen1996b}
\bibinfo{author}{Portengen, T.}, \bibinfo{author}{{\"O}streich, T.} \&
  \bibinfo{author}{Sham, L.~J.}
\newblock \bibinfo{title}{Theory of electronic ferroelectricity}.
\newblock \emph{\bibinfo{journal}{Phys. Rev. B}} \textbf{\bibinfo{volume}{54}},
  \bibinfo{pages}{17452} (\bibinfo{year}{1996}).

\bibitem{fei-cobden_2018}
\bibinfo{author}{Fei, Z.} \emph{et~al.}
\newblock \bibinfo{title}{Ferroelectric switching of a two-dimensional metal}.
\newblock \emph{\bibinfo{journal}{Nature}} \textbf{\bibinfo{volume}{560}},
  \bibinfo{pages}{336--339} (\bibinfo{year}{2018}).

\bibitem{Jia2017}
\bibinfo{author}{Jia, Z.-Y.} \emph{et~al.}
\newblock \bibinfo{title}{Direct visualization of a two-dimensional topological
  insulator in the single-layer
  $1{T}^{\ensuremath{'}}\ensuremath{-}\mathrm{WT}{\mathrm{e}}_{2}$}.
\newblock \emph{\bibinfo{journal}{Phys. Rev. B}} \textbf{\bibinfo{volume}{96}},
  \bibinfo{pages}{041108} (\bibinfo{year}{2017}).
\newblock \urlprefix\url{https://link.aps.org/doi/10.1103/PhysRevB.96.041108}.

\bibitem{Fei2017}
\bibinfo{author}{Fei, Z.} \emph{et~al.}
\newblock \bibinfo{title}{{Edge conduction in monolayer {WT}e$_2$}}.
\newblock \emph{\bibinfo{journal}{{Nature Physics}}}
  \textbf{\bibinfo{volume}{{13}}}, \bibinfo{pages}{677--682}
  (\bibinfo{year}{{2017}}).

\bibitem{Song2018}
\bibinfo{author}{Song, Y.-H.} \emph{et~al.}
\newblock \bibinfo{title}{Observation of {C}oulomb gap in the quantum spin
  {H}all candidate single-layer 1{T}$'$-{WT}e$_2$}.
\newblock \emph{\bibinfo{journal}{Nature Communications}}
  \textbf{\bibinfo{volume}{9}}, \bibinfo{pages}{4071} (\bibinfo{year}{2018}).
\newblock \urlprefix\url{https://doi.org/10.1038/s41467-018-06635-x}.

\bibitem{Sajadi2018}
\bibinfo{author}{Sajadi, E.} \emph{et~al.}
\newblock \bibinfo{title}{Gate-induced superconductivity in a monolayer
  topological insulator}.
\newblock \emph{\bibinfo{journal}{Science}} \textbf{\bibinfo{volume}{362}},
  \bibinfo{pages}{922--925} (\bibinfo{year}{2018}).
\newblock \urlprefix\url{https://science.sciencemag.org/content/362/6417/922}.
\newblock
  \eprint{https://science.sciencemag.org/content/362/6417/922.full.pdf}.

\bibitem{Varsano-Rontani2017}
\bibinfo{author}{Varsano, D.} \emph{et~al.}
\newblock \bibinfo{title}{{Carbon nanotubes as excitonic insulators}}.
\newblock \emph{\bibinfo{journal}{Nature Commun.}}
  \textbf{\bibinfo{volume}{8}}, \bibinfo{pages}{1461} (\bibinfo{year}{2017}).

\bibitem{Dean2017}
\bibinfo{author}{Li, J. I.~A.}, \bibinfo{author}{Taniguchi, T.},
  \bibinfo{author}{Watanabe, K.}, \bibinfo{author}{Hone, J.} \&
  \bibinfo{author}{Dean, C.~R.}
\newblock \bibinfo{title}{{Excitonic superfluid phase in double bilayer
  graphene}}.
\newblock \emph{\bibinfo{journal}{{Nature Phys.}}}
  \textbf{\bibinfo{volume}{{13}}}, \bibinfo{pages}{{751}}
  (\bibinfo{year}{{2017}}).

\bibitem{Kim2017}
\bibinfo{author}{Liu, X.}, \bibinfo{author}{Watanabe, K.},
  \bibinfo{author}{Taniguchi, T.}, \bibinfo{author}{Halperin, B.~I.} \&
  \bibinfo{author}{Kim, P.}
\newblock \bibinfo{title}{{Quantum Hall drag of exciton condensate in
  graphene}}.
\newblock \emph{\bibinfo{journal}{{Nature Phys.}}}
  \textbf{\bibinfo{volume}{{13}}}, \bibinfo{pages}{{746}}
  (\bibinfo{year}{{2017}}).

\bibitem{Kogar-Abbamonte_2017}
\bibinfo{author}{Kogar, A.} \emph{et~al.}
\newblock \bibinfo{title}{Signatures of exciton condensation in a transition
  metal dichalcogenide}.
\newblock \emph{\bibinfo{journal}{Science}} \textbf{\bibinfo{volume}{358}},
  \bibinfo{pages}{1314--1317} (\bibinfo{year}{2017}).

\bibitem{Kono2017}
\bibinfo{author}{Lu, Y.~F.} \emph{et~al.}
\newblock \bibinfo{title}{{Zero-gap semiconductor to excitonic insulator
  transition in Ta$_2$NiSe$_5$}}.
\newblock \emph{\bibinfo{journal}{{Nat. Commun.}}}
  \textbf{\bibinfo{volume}{{8}}}, \bibinfo{pages}{{14408}}
  (\bibinfo{year}{{2017}}).

\bibitem{Kaiser2018}
\bibinfo{author}{Werdehausen, D.} \emph{et~al.}
\newblock \bibinfo{title}{Coherent order parameter oscillations in the ground
  state of the excitonic insulator {T}a$_2${N}i{S}e$_5$}.
\newblock \emph{\bibinfo{journal}{Sci. Adv.}} \textbf{\bibinfo{volume}{4}},
  \bibinfo{pages}{aap8652} (\bibinfo{year}{2018}).

\bibitem{Noziers1964}
\bibinfo{author}{Nozi{\`e}rs, P.}
\newblock \emph{\bibinfo{title}{The theory of interacting {F}ermi systems}}
  (\bibinfo{publisher}{W. A. Benjamin Inc.}, \bibinfo{address}{New York},
  \bibinfo{year}{1964}).

\bibitem{Kohn1967b}
\bibinfo{author}{Kohn, W.}
\newblock \bibinfo{title}{Metals and insulators}.
\newblock In \bibinfo{editor}{de~Witt, C.} \& \bibinfo{editor}{Balian, R.}
  (eds.) \emph{\bibinfo{booktitle}{Many-body physics}},
  \bibinfo{pages}{351--411} (\bibinfo{publisher}{Gordon and Breach},
  \bibinfo{address}{New York}, \bibinfo{year}{1967}).

\bibitem{Cao2012}
\bibinfo{author}{Cao, T.} \emph{et~al.}
\newblock \bibinfo{title}{Valley-selective circular dichroism of monolayer
  molybdenum disulphide}.
\newblock \emph{\bibinfo{journal}{Nature Communications}}
  \textbf{\bibinfo{volume}{3}}, \bibinfo{pages}{1882} (\bibinfo{year}{2012}).
\newblock \urlprefix\url{https://doi.org/doi.org/10.1038/ncomms1882}.

\bibitem{Urbaszek2018}
\bibinfo{author}{Wang, G.} \emph{et~al.}
\newblock \bibinfo{title}{Colloquium: Excitons in atomically thin transition
  metal dichalcogenides}.
\newblock \emph{\bibinfo{journal}{Rev. Mod. Phys.}}
  \textbf{\bibinfo{volume}{90}}, \bibinfo{pages}{021001}
  (\bibinfo{year}{2018}).
\newblock
  \urlprefix\url{https://link.aps.org/doi/10.1103/RevModPhys.90.021001}.

\bibitem{Yafet1963}
\bibinfo{author}{Yafet, Y.}
\newblock \bibinfo{title}{$g$ factors and spin-lattice relaxation of conduction
  electrons}.
\newblock In \bibinfo{editor}{Seitz, F.} \& \bibinfo{editor}{Turnbull, D.}
  (eds.) \emph{\bibinfo{booktitle}{Solid state physics}},
  vol.~\bibinfo{volume}{14}, \bibinfo{pages}{1--98}
  (\bibinfo{publisher}{Academic Press}, \bibinfo{address}{New York},
  \bibinfo{year}{1963}).

\bibitem{Fatemi2018}
\bibinfo{author}{Fatemi, V.} \emph{et~al.}
\newblock \bibinfo{title}{Electrically tunable low-density superconductivity in
  a monolayer topological insulator}.
\newblock \emph{\bibinfo{journal}{Science}} \textbf{\bibinfo{volume}{362}},
  \bibinfo{pages}{926--929} (\bibinfo{year}{2018}).
\newblock \urlprefix\url{https://science.sciencemag.org/content/362/6417/926}.
\newblock
  \eprint{https://science.sciencemag.org/content/362/6417/926.full.pdf}.

\bibitem{Giannozzi2009}
\bibinfo{author}{Giannozzi, P.} \emph{et~al.}
\newblock \bibinfo{title}{Quantum {ESPRESSO}: a modular and open-source
  software project for quantum simulations of materials}.
\newblock \emph{\bibinfo{journal}{J. Phys.: Condens. Matter}}
  \textbf{\bibinfo{volume}{21}}, \bibinfo{pages}{395502}
  (\bibinfo{year}{2009}).

\bibitem{perdew1996generalized}
\bibinfo{author}{Perdew, J.~P.}, \bibinfo{author}{Burke, K.} \&
  \bibinfo{author}{Ernzerhof, M.}
\newblock \bibinfo{title}{Generalized gradient approximation made simple}.
\newblock \emph{\bibinfo{journal}{Phys. Rev. Lett.}}
  \textbf{\bibinfo{volume}{77}}, \bibinfo{pages}{3865} (\bibinfo{year}{1996}).

\bibitem{Marini2009}
\bibinfo{author}{Marini, A.}, \bibinfo{author}{Hogan, C.},
  \bibinfo{author}{Gr{\"u}ning, M.} \& \bibinfo{author}{Varsano, D.}
\newblock \bibinfo{title}{Yambo: {A}n ab initio tool for excited state
  calculations}.
\newblock \emph{\bibinfo{journal}{Comput. Phys. Commun.}}
  \textbf{\bibinfo{volume}{180}}, \bibinfo{pages}{1392--1403}
  (\bibinfo{year}{2009}).

\bibitem{sangalli2019}
\bibinfo{author}{Sangalli, D.} \emph{et~al.}
\newblock \bibinfo{title}{Many-body perturbation theory calculations using the
  yambo code}.
\newblock \emph{\bibinfo{journal}{Journal of Physics: Condensed Matter}}
  (\bibinfo{year}{2019}).
\newblock \urlprefix\url{http://iopscience.iop.org/10.1088/1361-648X/ab15d0}.

\bibitem{Godby1989}
\bibinfo{author}{Godby, R.~W.} \& \bibinfo{author}{Needs, R.~J.}
\newblock \bibinfo{title}{Metal-insulator transition in {K}ohn-{S}ham theory
  and quasiparticle theory}.
\newblock \emph{\bibinfo{journal}{Phys. Rev. Lett.}}
  \textbf{\bibinfo{volume}{62}}, \bibinfo{pages}{1169--1172}
  (\bibinfo{year}{1989}).

\bibitem{Rozzi2006}
\bibinfo{author}{Rozzi, C.~A.}, \bibinfo{author}{Varsano, D.},
  \bibinfo{author}{Marini, A.}, \bibinfo{author}{Gross, E. K.~U.} \&
  \bibinfo{author}{Rubio, A.}
\newblock \bibinfo{title}{Exact {C}oulomb cutoff technique for supercell
  calculations}.
\newblock \emph{\bibinfo{journal}{Phys. Rev. B}} \textbf{\bibinfo{volume}{73}},
  \bibinfo{pages}{205119} (\bibinfo{year}{2006}).
\newblock \urlprefix\url{https://link.aps.org/doi/10.1103/PhysRevB.73.205119}.

\bibitem{kan2014structures}
\bibinfo{author}{Kan, M.} \emph{et~al.}
\newblock \bibinfo{title}{Structures and phase transition of a {M}o{S}$_2$
  monolayer}.
\newblock \emph{\bibinfo{journal}{The Journal of Physical Chemistry C}}
  \textbf{\bibinfo{volume}{118}}, \bibinfo{pages}{1515--1522}
  (\bibinfo{year}{2014}).

\bibitem{LuttingerKohn1955}
\bibinfo{author}{Luttinger, J.~M.} \& \bibinfo{author}{Kohn, W.}
\newblock \bibinfo{title}{Motion of electrons and holes in perturbed periodic
  fields}.
\newblock \emph{\bibinfo{journal}{Phys. Rev.}} \textbf{\bibinfo{volume}{97}},
  \bibinfo{pages}{869} (\bibinfo{year}{1955}).
\newblock \urlprefix\url{https://link.aps.org/doi/10.1103/PhysRev.97.869}.

\bibitem{Cudazzo2011}
\bibinfo{author}{Cudazzo, P.}, \bibinfo{author}{Tokatly, I.~V.} \&
  \bibinfo{author}{Rubio, A.}
\newblock \bibinfo{title}{Dielectric screening in two-dimensional insulators:
  {I}mplications for excitonic and impurity states in graphane}.
\newblock \emph{\bibinfo{journal}{Phys. Rev. B}} \textbf{\bibinfo{volume}{84}},
  \bibinfo{pages}{085406} (\bibinfo{year}{2011}).

\bibitem{Kane2005b}
\bibinfo{author}{Kane, C.~L.} \& \bibinfo{author}{Mele, E.~J.}
\newblock \bibinfo{title}{${Z}_{2}$ topological order and the quantum spin
  {H}all effect}.
\newblock \emph{\bibinfo{journal}{Phys. Rev. Lett.}}
  \textbf{\bibinfo{volume}{95}}, \bibinfo{pages}{146802}
  (\bibinfo{year}{2005}).
\newblock
  \urlprefix\url{https://link.aps.org/doi/10.1103/PhysRevLett.95.146802}.

\end{thebibliography}



\begin{addendum}
 
 \item We thank Cl\'audia Cardoso for the rendering of figures 1a, 2a,
 and 4d. M.P.~thanks Giancarlo Cicero for illuminating discussions
 in the early stages of this project and acknowledges Tor Vergata University for financial support through the mission sustainability
project 2DUTOPI.

 This work was supported in part by the 
 European Union H2020-INFRAEDI-2018-1 programme under grant agreement No. 824143 project 453 “MaX - materials design at the exascale”.
 This work was also supported by MIUR-PRIN2017 No.~2017BZPKSZ `Excitonic insulator in two-dimensional long-range interacting systems (EXC-INS)'. The authors acknowledge PRACE for awarding them access to the Marconi system based in Italy at CINECA.
 \item[Author contributions] D.V.~and M.P.~developed the first-principles many-body perturbation theory calculations and analysis, M.R.~developed the 
 self-consistent mean-field model and wrote the paper, all authors initiated 
 this project, contributed to the analysis of data, and critically
 discussed the paper.
 \item[Supplementary Information] is available in the online
 version of the paper.
 \item[Competing Interests] The authors declare that they have no
 competing financial interests.
 \item[Correspondence] Correspondence and requests for materials
 should be addressed to M.R.~(email: massimo.rontani@nano.cnr.it).
\end{addendum}


\end{document}